# Millimeter-Wave Integrated Silicon Devices:
# Active versus Passive — The Eternal Struggle Between Good and Evil

*(Invited Paper)*

**Michele Spasaro and Domenico Zito**
Department of Engineering - Electrical and Computer Engineering, Aarhus University, Finlandsgade 22, 8200 Aarhus N, Denmark

*Abstract*— With the extreme scaling, active devices in both CMOS and BiCMOS technologies have reached outstanding $f_T/f_{max}$, enabling an ever-increasing number of existing and emerging applications in the microwave and millimeter wave (mm-wave) frequency range. The increase in transistors $f_T/f_{max}$ has been so much significant that the performance of microwave and mm-wave ICs are limited mainly by losses in passive devices.

In this paper, we address a discussion on qualitative and quantitative aspects that may help to further unveil the impact of such losses on the overall circuit performance and stimulate the adoption of effective loss-aware design methodologies. As example, we report the results related to the design of low power mm-wave low noise amplifiers (LNAs). Our results show how, in low power regime, the performances of mm-wave LNAs are dominated by losses in passive components. We also show how loss-aware design methodologies can mitigate the performance degradation due to passives, resulting as an important tool to get the full potential out of the active devices available today.

## 1. Introduction

With reference to the silicon technology evolution, fully depleted technologies have become predominant in semiconductor industry with the advent of the 28nm CMOS technology node [1]. Two main types of fully depleted active devices have emerged: FinFET CMOS and Fully Depleted Silicon on Insulator (FD-SOI) CMOS devices.

The most advanced technology nodes in mass production in main foundries worldwide are based on FinFET devices. This is the case of TSMC's 7nm logic technology [2], GlobalFoundries' 14LPP 14nm FinFET technology [3], and Samsung's 7LPP 7nm technology [4], to name a few.

The research conducted by CEA-Leti together with STMicroelectronics led to the first FD-SOI technology in mass production [5]. After the introduction of world-first 28nm FD-SOI technology by STMicroelectronics, GlobalFoundries and Samsung have also included FD-SOI devices in their portfolios: GlobalFoundries offers 22FDX®, a 22nm FD-SOI technology [6]. Samsung is mass producing 28FDS, a 28nm FD-SOI process [7].

Despite the latest CMOS technology nodes are approaching the performances of heterojunction bipolar transistors (HBTs) in BiCMOS technologies, the latter have still some advantages primarily related to the more mature back-end-of-line (BEOL), which may lead typically to better passive devices, lower sensitivity of the intrinsic transistor $f_T$ to parasitic capacitances of metallization layers, higher reliability and lower costs [8, 9].

With the extreme scaling down to a few nanometers, so approaching the fundamental limits of integration (before jumping into the sub-atomic scale electronics era), active devices have reached outstanding performances, enabling an ever-increasing range of microwave and millimeter wave (mm-wave) applications, such as high data-rate communications for next generation wireless networks (5G and beyond), virtual reality, vehicular networks, radio-astronomy, Earth observations and satellite communications [10, 11].

Typical figures of merit adopted to evaluate the high-frequency capability of technologies

This work was supported in part by the European Commission through the European H2020 FET OPEN project IQubits (Grant Agreement number 829005), and in part by the Faculty of Science and Technology, Aarhus University, and the Poul Due Jensen Foundation.

are the transition (or cut-off) frequency ($f_T$) and maximum oscillation frequency ($f_{max}$) of the transistors. For CMOS technologies, the typical current densities for peak−$f_T$ and peak−$f_{max}$ are equal approximately to 0.3 mA/μm and 0.2 mA/μm, respectively [12]. These were proved to be largely invariant across technology nodes and operating frequencies, and are today largely adopted in most of the state-of-the-art solutions for radio-frequency (microwaves and mm-waves) IC design.

Despite the higher level of planarity of FD-SOI technology may lead to some competitive advantages with respect to FinFET technology, both CMOS and BiCMOS technologies have reached outstanding $f_T$/$f_{max}$. For example, 22FDX® includes thin-oxide n-FET devices with measured $f_T$/$f_{max}$ as high as 347/371 GHz [6, 13]. In the same studies, n-FETs in 14nm FinFET technology by GlobalFoundries feature $f_T$/$f_{max}$ of 314/180 GHz, revealing the superior radio-frequency (RF) performance of FD-SOI technologies. Also, the 28nm FD-SOI CMOS technology by STMicroelectronics features maximum $f_T$/$f_{max}$ higher than 300 GHz [1].

As for BiCMOS technologies, in the 90nm SiGe:C BiCMOS technology by NXP semiconductors, the HBTs feature $f_T$/$f_{max}$ ~230/400 GHz [14]. SG13S, a 0.13um SiGe:C BiCMOS technology by IHP, features npn-HBTs with maximum $f_T$/$f_{max}$ of 250/340 GHz [15]. BiCMOS055, the 55nm SiGe BiCMOS technology by STMicroelectronics, features HBTs with maximum $f_T$/$f_{max}$ of 320/370 GHz [9].

The extreme scaling, also supported by atomic-layer deposition techniques, have led to stunning improvements of transistor performances, which have enabled massive integration of innovative mm-wave solutions both for existing and emerging applications.

These improvements have been so much significant that the transistors have allowed overcoming the dramatic high-frequency performance degradation of passive devices such as spiral inductors and transformers, metal-oxide-metal (MOM) and metal-insulator-metal (MIM) capacitors, transmission lines, and then enabling the implementation of complex integrated systems.

Moving forward and considering the technology progression towards higher $f_T$/$f_{max}$ as fundamental requirements for operations in the terahertz range above the mm-waves, i.e. above 300 GHz, one could argue about what will be the actual capability of future transistors to overcome the even higher performance degradation due to passive components at those frequencies.

Most of the technology efforts are addressed to active devices, but despite the expected improvements, the severe limitations of passive devices lead regularly to the same conclusion: the need of more effective design methodologies, circuit topologies and system architectures. These are definitively strategic and legitim ambitions that should be pursued insistently, however they go beyond the technology evolution.

As the losses in passive devices introduce significant roadblocks to the performance of microwave and mm-wave ICs, in this paper we address a discussion on qualitative and quantitative aspects that may help to further unveil their impact on the overall circuit performance, and allow developing further awareness and stimulate the adoption of effective loss-aware design strategies.

To do this, here we consider as reference the case of low noise amplifiers (LNAs), which are among the most critical building blocks for the overall system performance, as they must fulfill a quite large set of diverse requirements, such as gain, noise, bandwidth, impedance matching, stability, linearity, and power consumption.

Most of the theoretical approaches to LNA design, such as [16], are based on lossless matching networks (MNs). Such approaches are developed with the objective to exploit the full potential of active devices. For example, [16] targets operation near $f_{max}$. Nonetheless, the theoretical bounds predicted by such theories are far to be approached because of the unavoidable losses in real MNs [17, 18].

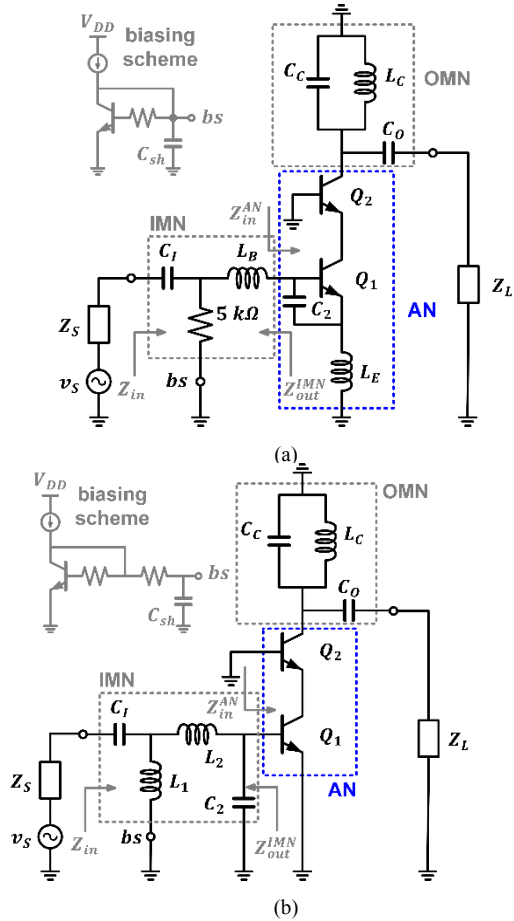

**Fig. 1.** Small-signal analysis circuits. The inserts (in grey) refer to examples of possible bias circuitry, not considered in our studies. (a) Cascode amplifier with inductive degeneration $L_E$ and additional base-emitter capacitor $C_2$. (b) LNA circuit topology with a four-element IMN and no feedback MN. The blue enclosures identify the active networks (ANs).

Ideal matching networks do not exist in practice, even less on silicon substrates, thereby it is interesting to focus our attention on how to design high-frequency circuits by taking into account also the losses of passives.

Recently our research group has developed the *theory of special noise invariants* [19], which has led to a novel loss-aware LNA design methodology. We will reuse the concepts reported therein in order to quantify the dramatic impact of losses and the benefits that can derive from adopting loss-aware design approaches in order to mitigate the performance degradation due to passives.

The paper is organized as follows. In Section 2, we recall some fundamentals. In particular, we focus on the impact of losses in MNs on the noise performance of low-power tuned mm-wave LNAs in a commercial BiCMOS technology. We report and discuss the qualitative and quantitative results of our study, in order to appreciate the potential of loss-aware design methodologies. In Section 3, we draw the conclusions.

## 2. Low-Power Tuned mm-Wave LNA

The cascode amplifier with inductive degeneration, as in Fig. 1(a), is the most widespread circuit topology for tuned LNAs [20, 21]. This topology is typically tailored to a design methodology which considers inductors and capacitors as lossless passives, as recalled hereinafter (see Section 2.C).

Recently, we proposed the topology of Fig. 1(b) [19], in which the active network (AN) consists of the cascode amplifier, without the feedback elements $L_E$ and $C_2$. The input matching network (IMN) comprises two capacitors and two integrated inductors, and is tailored to a loss-aware design methodology presented in [19] and recalled hereinafter (see Section 2.D).

In this paper, we compare the two circuit topologies, each with the own design methodology, with the objective to derive qualitative and quantitative insights about the impact of losses caused by the passive devices, such as inductors and capacitors.

Hereinafter we recall some theoretical results emerged from our studies. We discuss the assumptions underlying the comparative study carried out in this paper for low power regime and show that the LNA performances are dominated by matching network losses, with heavy impact on the LNA of Fig.1(a), whereas the impact of losses is mitigated in the LNA of Fig. 1(b), whose design methodology allows achieving superior gain and negligible noise figure degradation.

### A. Recalls on Noise Invariants

The characterization of noise in active devices considered as a two-port AN is often summarized by four noise parameters [22]: the optimum noise impedance $(Z_{ON}^{AN})$, the equivalent noise resistance $(R_n^{AN})$, and the minimum equivalent noise temperature ($T_{Nmin}^{AN}$).

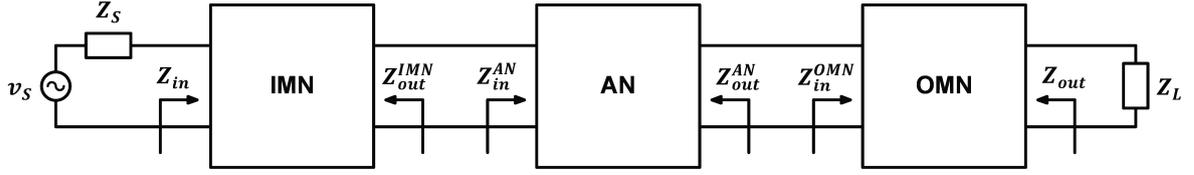

**Fig. 2.** Two-port amplifier consisting of a two-port linear noisy active network (AN) embedded in a matching network (MN) comprising an input matching network (IMN) and an output matching network (OMN). The notation for the output and input impedances of each stage is also introduced in the figure.

Recently, the *theory of special noise invariants* [19] has shed a new light on noise parameters.

$T_{Nmin}^{AN}$ is the minimum value of the equivalent noise temperature ($T_N$) for any LNA comprising AN and lossless reciprocal input and output feedforward MNs, as shown in Fig. 2. These MNs preserve $T_{Nmin}^{AN}$, i.e., $T_{Nmin}$ of the LNA is equal to $T_{Nmin}^{AN}$, and thereby $T_{Nmin}^{AN}$ is a *special invariant*. As invariant, $T_{Nmin}^{AN}$ is a valuable tool to study the impact of losses in MNs.

MNs including feedback paths do not preserve $T_{Nmin}^{AN}$ of any AN [19]. For some MNs, $T_{Nmin}$ is even lower than $T_{Nmin}^{AN}$. Inductive degeneration, as shown in Fig. 3, is an example of feedback transformation lowering the minimum equivalent noise temperature [19]. Since $T_{Nmin}$ can be lower than $T_{Nmin}^{AN}$, the latter is a lower bound to $T_N$ only if the MN does not include feedback paths.

Another quantity characterizing the noise performance of an LNA is the *power-added noise temperature ($MT_0$)* [23, 24].

For the LNA of Fig. 4, with a two-port AN embedded in a four-port MN comprising feedback paths and/or feedforward MNs, $MT_0$ is greater than the AN *minimum power-added noise temperature* $(MT_0)_{min}^{AN}$, which is a characteristic quantity of the active network [23, 24].

Moreover, $(MT_0)_{min}$ of the LNA is equal to $(MT_0)_{min}^{AN}$, if MN is lossless. This means that $(MT_0)_{min}^{AN}$ is a *general invariant*. Also, since $MT_0$ is related to $T_N$ and the available gain $G_A$ of the LNA as follows

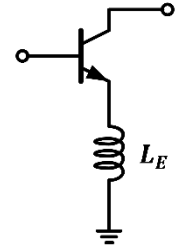

**Fig. 3.** Common-emitter stage with inductive degeneration.

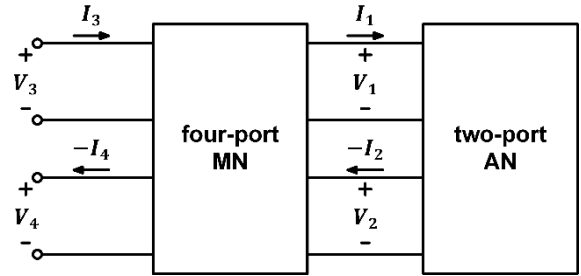

**Fig. 4.** Embedding of a two-port AN in a four-port lossless MN. The transformation produces a new two-port network.

$$MT_0 = T_N/(1 - 1/G_A), \qquad (1)$$

it is a figure of merit which allows comparing gain and noise performances of two LNAs.

### B. Assumptions

In our comparative study, the LNAs are designed in a 0.13μm SiGe:C BiCMOS technology commercially available.

The transistors have the same size and current density (i.e. for minimum noise figure of the cascode amplifier), for both LNAs.

For both LNAs, the output coupling capacitor $C_O$ and the $C_C$-$L_C$ tank form the output matching network (OMN), designed for maximum power transfer. The two LNAs have equal values of $C_O$, $C_C$, and $L_C$. The transistors have minimum size in order to minimize the power consumption, which amounts to 760 μW, from a supply voltage of 2 V.

All passive components are considered lossy, unless stated otherwise. All capacitors consist of metal-insulator-metal (MIM) capacitors available within the design kit, which includes quite accurate models. MIM capacitors feature a good area efficiency and exhibit capacitances larger than few femtofarads [25]. Metal-oxide-metal (MOM) capacitors allow smaller capacitances; however, capacitances smaller than few femtofarads may be required rarely in LNA designs. The quality factor (Q) of the spiral inductors follow roughly a linear regression between 25 for 50 pH and 15 for 500 pH. The typical quality factor of spiral inductors and MIM capacitors are reported in Fig. 5.

### C. Impact of Losses on LNA Performances

Here, we address the design of the low power LNAs operating at 60 GHz.

First, we consider the cascode amplifier with inductive degeneration and additional base-emitter capacitor $C_2$ of Fig. 1(a). We will show that, in low power regime, losses in matching networks dominate the noise performance of the LNA. As illustrated in Fig. 1(a), the LNA is modeled as a two-port AN connected to feedforward input and output MNs. The AN comprises the transistors $Q_1$ and $Q_2$ in cascode configuration, the emitter degeneration inductor $L_E$ and the additional capacitor $C_2$.

The two transistors are equally sized and biased with the collector current density $J$ that minimizes the minimum equivalent noise temperature $T_{Nmin}^{CAS}$ of the cascode amplifier.

Fig. 6 plots $T_{Nmin}^{CAS}$ as a function of $J$. The minimum value of $T_{Nmin}^{CAS}$ amounts to 276 K and is achieved for $J$ equal to 6.6 mA/μm². The corresponding minimum noise figure $NF_{min}^{CAS}$ is equal to 2.9 dB.

In ideal terms, $C_2$, $L_E$ and $L_B$ perform a sequence of lossless circuit transformations, according to the design methodology

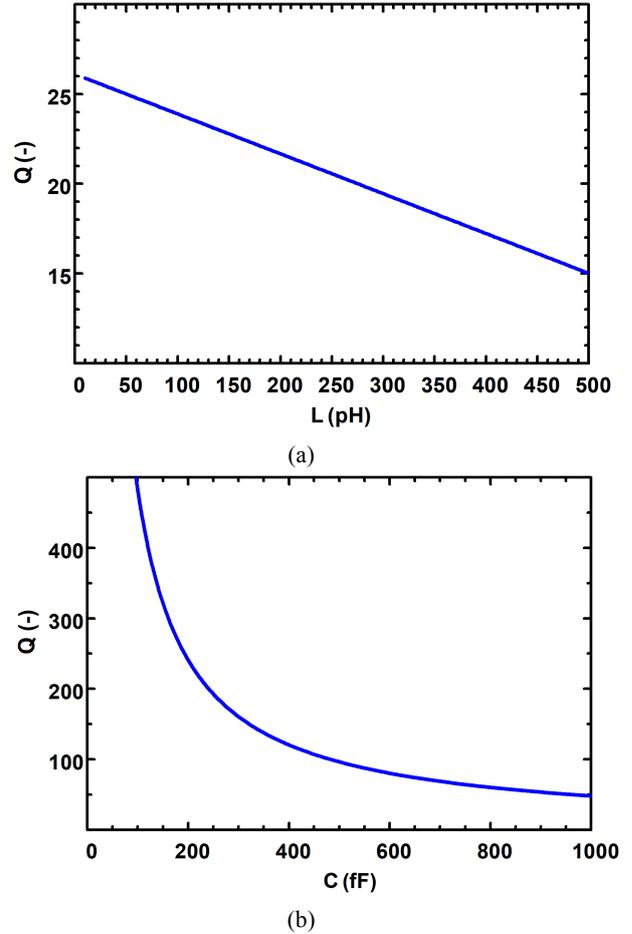

**Fig. 5.** (a) Quality factor of the inductors. (b) Quality factor of MIM capacitors as intrinsic devices.

developed in [20] and formalized for any LNA topology in [19] as follows.

1) The lossless feedback transformation sets AN optimum noise impedance $Z_{ON}^{AN}$ equal to the complex conjugate of AN input impedance $Z_{in}^{AN}$:

$$Z_{ON}^{AN} = \left(Z_{in}^{AN}\right)^* \qquad (2)$$

2) The lossless IMN sets LNA input impedance $Z_{in}$ equal to the complex conjugate of the source impedance $Z_S$:

$$Z_{in} = Z_S^*. \qquad (3)$$

As a lossless reciprocal two-port network has equal reflection coefficients at its own two ports, we

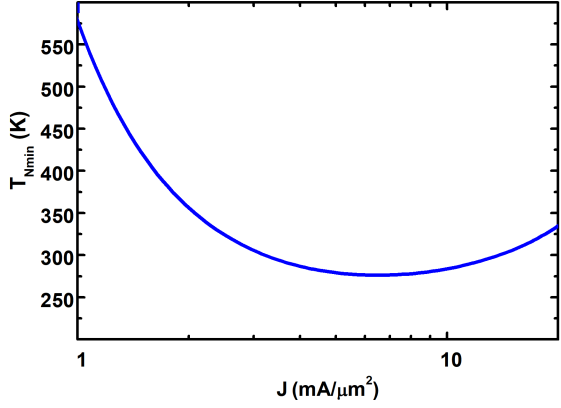

**Fig. 6.** $T_{Nmin}^{CAS}$ of the cascode amplifier as a function of the current density $J$.

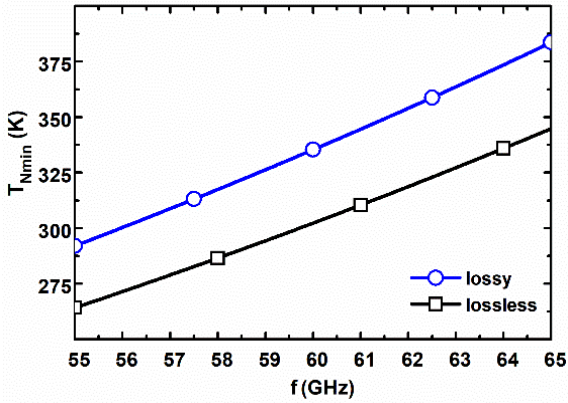

**Fig. 7.** $T_{Nmin}^{AN}$ of the AN. The green and blue curves refer to the cases of lossy and lossless passives, respectively.

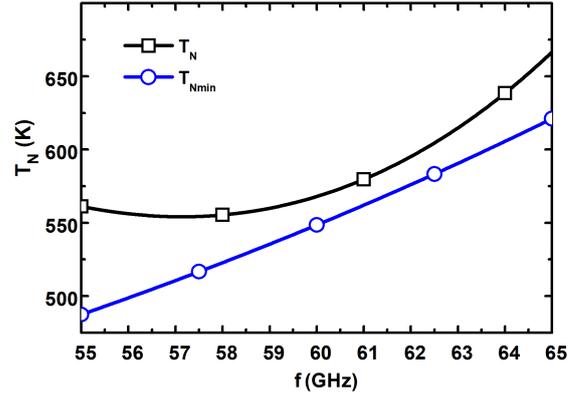

**Fig. 8.** $T_N$ and $T_{Nmin}$ of the LNA with inductive degeneration.

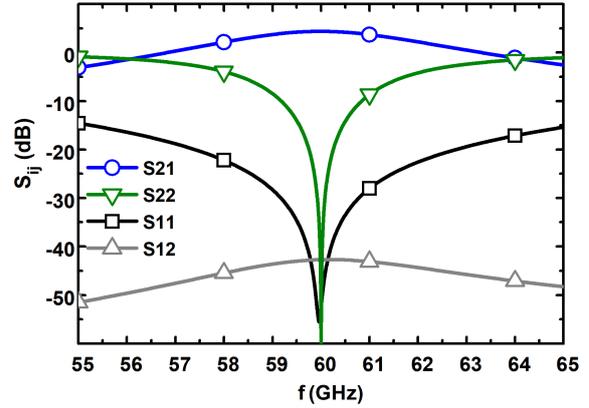

**Fig. 9.** S-parameters of the LNA with inductive degeneration.

have that $Z_{out}^{IMN} = \left(Z_{in}^{AN}\right)^*$, which once combined with (2) gives:

$$Z_{out}^{IMN} = Z_{ON}^{AN}. \quad (4)$$

We refer to the design methodology above as *input integrated matching for maximum power transfer and minimum noise figure* (IIM for MPmN) [26, 19].

IIM for MPmN is a design methodology applicable to any LNA topology. In the specific case of the LNA topology of Fig. 1(b), the design methodology becomes as follows. The additional capacitor $C_2$ [21] lowers the optimum-noise resistance $Re\{Z_{ON}^{AN}\}$ and is adopted to achieve condition (2) without resorting to large transistors. The emitter degeneration inductor $L_E$ raises [20] the input resistance of the cascode amplifier and the base inductor $L_B$ resonates out the input reactance, enabling the achievement of (3).

The performance of the LNA and the impact of MN losses are analyzed through Figs. 7-10.

Fig. 7 shows $T_{Nmin}^{AN}$ for the cases of lossy and lossless passive components $L_E$ and $C_2$. Note that, because of the losses in $L_E$ and $C_2$, $T_{Nmin}^{AN}$ raises from 303 K (for lossless $L_E$ and $C_2$) to 336 K.

Fig. 8 shows $T_N$ and $T_{Nmin}$ of the entire LNA, which features impedance matching for minimum noise temperature. However, note that the losses in passive components raise $T_{Nmin}$ to 548 K, i.e. a 98.5 % increase with respect to $T_{Nmin}^{CAS}$.

Fig. 9 shows the S parameters of the LNA. The losses in passive components have also a detrimental impact on LNA gain: $S_{21}$ amounts to 4.4 dB, despite the maximum power transfer is achieved at both LNA ports and AN exhibits a maximum stable gain $G_{MSG}^{AN}$ of 24.4 dB, as shown in Fig. 10.

Despite the excellent performance of the active devices ($T_{Nmin}^{CAS}$ of 276 K and $G_{MSG}^{AN}$ of 24.4 dB) the final LNA features $T_N$ of 568 K (i.e. more than the double of $T_{Nmin}^{CAS}$) and S$_{21}$ of only 4.4 dB (i.e. 20 dB lower than $G_{MSG}^{AN}$). This dramatic performance degradation is only due to MN losses.

This shows how, in low power and high-frequency applications, the performance of the LNA is dominated by the losses in passive components, which impede the exploitation of the full potential of the transistors. Also, this means that traditional design approaches based on lossless matching networks is insufficient to get the full potential out of the active devices, and then this leads to investigate more advanced loss-aware design methodologies, which are key to achieve acceptable performances, as shown hereinafter.

### D. LNA Design through Loss-Aware Design Methodology

As discussed above, the lossy MN consisting of $C_2$-$L_E$ has a detrimental impact on the performance of the LNA of Fig. 1(a). For this reason, the AN of the topology of Fig. 1(b) adopts a cascode amplifier without any feedback MN. Moreover, in order to optimize the LNA performance, we propose a design methodology that takes into account also the losses of the IMN, as discussed below.

Because of the losses, the reflection coefficients at the two ports of IMN are unequal, in general. If the IMN is designed so as to achieve maximum power transfer at the input port of the LNA, part of the signal power incident on AN input port is reflected back toward IMN. The shunt inductor $L_1$ and capacitor $C_2$ allow exploiting the losses in order to vary the IMN output reflection coefficient $\Gamma_{out}^{IMN}$ (or IMN output impedance $Z_{out}^{IMN}$), while keeping the input reflection coefficient as low as possible.

In [19] we showed that a trade-off exists in the design of lossy IMNs. In order to minimize the AN noise contribution, $Z_{out}^{IMN}$ must be equal to $Z_{ON}^{AN}$. In order to minimize IMN noise contribution, the IMN available gain $G_A^{IMN}$ must be the largest possible. Unfortunately, $G_A^{IMN}$ and $Z_{out}^{IMN}$ cannot be chosen independently for a given IMN topology. As a consequence, a trade-off between the noise contributions of IMN and AN is needed to minimize the noise figure of the cascade of the two networks.

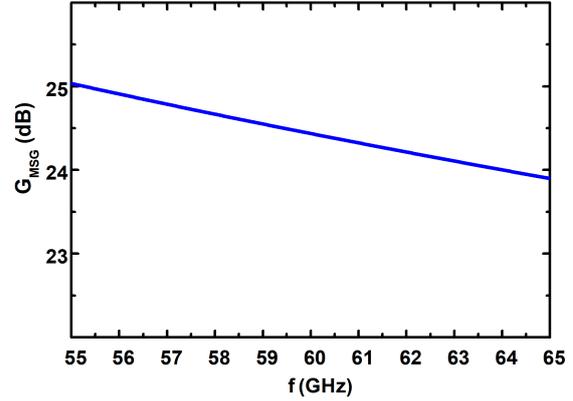

**Fig. 10.** Maximum stable gain of the AN of the amplifier with inductive degeneration.

To sum up, IMN losses are exploited effectively if the IMN is designed such that it maximizes power transfer at the LNA input port and minimizes the equivalent noise temperature of the cascade of IMN and AN, simultaneously. This condition is referred to as *input integrated matching for maximum power transfer and minimum cascade noise* (IIM for MPmCN). This can be achieved by sizing properly the lossy IMN. An IMN comprising both capacitors and integrated inductors can dissipate significant amounts of input power [27]. If the IMN comprises at least three components, as a general guideline, two can be sized so as to achieve maximum power transfer and the third one can be exploited to minimize the cascade noise.

The IMN of the LNA of Fig. 1(b) comprises two capacitors and two integrated inductors, so that the IIM for MPmCN can be applied to LNA design. By applying this design methodology, the LNA of Fig. 1(b) achieves a superior gain S$_{21}$ of 8.4 dB with a slightly degraded $T_N$ of 703 K (*NF* of 5.3 dB), as shown in Fig. 11. Maximum power transfer is achieved at both LNA ports, as shown in Fig. 12.

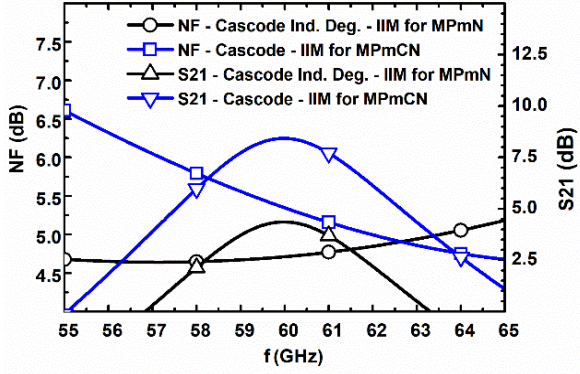

**Fig. 11.** Transducer gain $S_{21}$ and noise figure NF of the two LNAs.

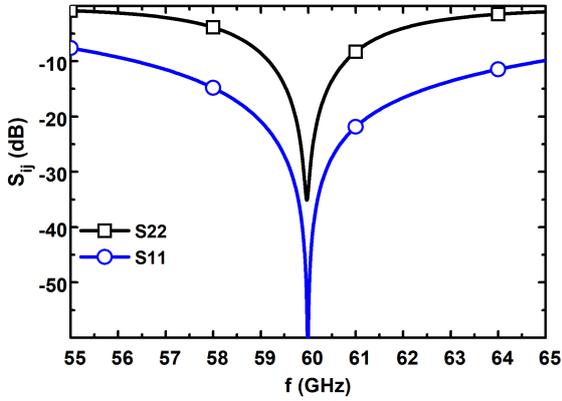

**Fig. 12.** $S_{11}$ and $S_{22}$ of the cascode LNA of Fig. 1(b).

For both LNAs, the geometrically derived stability factor $\mu$ is greater than one at all frequencies.

Table 1 reports the circuit sizing of the two LNAs. Table 2 summarizes the performance. The LNA circuit of Fig. 1(b) exhibits a 4 dB higher gain with a negligible 0.6 dB NF degradation and thereby a lower $MT_0$ (about 76 K lower) with respect to the cascode LNA with inductive degeneration of Fig. 1(a).

The IMN of the LNAs of Fig.1(a) and (b) have an available gain of -0.4 dB and -1.1 dB, respectively. However, the available gain of AN plus OMN is equal to 4.8 for the LNA of Fig. 1(a) and 9.5 for the LNA of Fig. 1(b). The worse performance of the lossy IMN in the LNA of Fig. 1(b) is compensated by the higher gain of the AN, since no feedback element reduces it, enabling better exploitation of the gain potential of active devices.

Still, for the LNA of Fig. 1(b), $T_N$ is more than the double of $T_{Nmin}^{CAS}$ and the full potential of active devices is far to be approached.

**Table 1.** Circuit sizing

| Topology | $C_I$ | $L_E/L_1$ | $C_2$ | $L_B/L_2$ | $L_C$ | $C_C$ | $C_O$ |
|---|---|---|---|---|---|---|---|
| Cascode Ind. Deg. | 1 pF | 113 pH | 16 fF | 241 pH | 73 pH | 81 fF | 15 fF |
| Cascode | 134 fF | 85 pH | 15 fF | 222 pH | 73 pH | 81 fF | 15 fF |

**Table 2.** Comparison between the two LNAs

| Topology | $S_{21}$ | $NF$ | $MT_0$ | $iP_{1dB}$ | $\epsilon_{GN}$ |
|---|---|---|---|---|---|
| Cascode Ind. Deg. | 4.4 dB | 4.7 dB | 897 K | -9 dBm | -0.3 dB |
| Cascode | 8.4 dB | 5.3 dB | **821 K** | -13 dBm | **3.1 dB** |

In order to compare the performance of the two LNAs, a further figure of merit, the *gain-noise excess* ($\epsilon_{GN}$), is defined as follows

$$\epsilon_{GN} = S_{21}[dB] - NF. \qquad (5)$$

The $\epsilon_{GN}$ can be considered as a qualitative reference for preliminary evaluations of the essential performances of the LNA, in relation to the potential benefits to the receiver performances. The LNAs of Fig. 1(a) and (b) show $\epsilon_{GN}$ of -0.3 dB and 3.1 dB, respectively. This means that the LNA of Fig. 1(a), with a negative $\epsilon_{GN}$ (NF is higher than the gain), could be detrimental for the overall receiver performances and its adoption could be unjustified in terms of cost-benefits, as this depends on the performances of the next stages of the receiver and could even result in performance worsening and power wasting. As for the LNA of Fig. 1(b), the $\epsilon_{GN}$ exceeds slightly 3dB, which could be considered as a preliminary qualitative requirement for its adoption in the receiver. Both $MT_0$ and $\epsilon_{GN}$ support the superiority of the cascode amplifier featuring IIM for MPmCN, despite the slight increase in noise figure.

The IIM for MPmCN emerges as effective, demonstrating that design methodologies accounting for MN losses are key to improve the performances.

## 3. Conclusions

In this paper we discussed on the evolution of silicon technologies that have led to transistors with excellent performance for operations at the millimeter waves, with predictions on $f_T/f_{max}$ of future technology nodes approaching the terahertz.

Despite the stunning progression of active devices, the losses in passive devices introduce significant limitations to reach the full potential of the available transistors in mm-wave IC design.

Beyond more general qualitative aspects, we focused our discussion on some quantitative aspects of the mm-wave LNA performance degradation with the objective of unveiling their impact on the overall circuit performance and stimulate the adoption of effective loss-aware design strategies.

The results show the dramatic impact of losses of passive components on overall circuit performance. Above all, they show how the traditional design approaches based on lossless matching network may fail in arranging a competitive design, leaving the potential of technology unexploited. Unlike these, we show how the design methodology taking into account losses allows getting the full potential out of it, leading to performance that would not be achieved otherwise.

In other terms, the results indicate the strong need of further technology advances towards better passives, likely with the introduction of new materials and technology processes. Examples of these could be materials with ferroelectric, ferromagnetic properties, superior insulator properties, which would be key for future emerging high-frequency applications, including spin readout and manipulation circuits for quantum computing, today among the most far-reaching and challenging information and communication technologies.

Moving forward, the extreme scaling down to a few nanometers will enable better and better active devices, but better passives are strongly required to get the full potential out of them. Without advances on passives, the potential of the entire technology progression, primarily focused on active devices, could be dramatically impaired and unexploited. Advanced lithography and atomic-layer deposition techniques, together with the introduction of new materials, could play a significant role, as the reduction of form factor of passives that is key to reduce the losses and achieve better quality factor.

Last, but not least, loss-aware design methodologies are a must, and key to design integrated circuits and systems in the mm-wave frequency range and beyond, and thereby instrumentals in this "eternal" struggle between good and evil.

**Acknowledgments.** The authors are grateful to Keysight Technologies for their support through the donation of equipment and cad tools.

**References**

[1] A. Cathelin, "*Fully depleted silicon on insulator devices CMOS: the 28-nm node is the perfect technology for analog, RF, mmW, and mixed-signal system-on-chip integration*", IEEE Solid-State Circuits Mag., **9**(4), pp. 18-26, 2017.

[2] Taiwan Semiconductor Manufacturing Company Limited, "*Logic Technology*", [Online]. Available: https://www.tsmc.com/english/dedicatedFoundry/technology/logic.htm, Accessed on Jul. 2019.

[3] GlobalFoundries, "Product brief: 14LPP FinFET technology". [Online]. Available: https://www.globalfoundries.com/technology-solutions/cmos/performance/14lpp, Accessed on Jul. 2019.

[4] Y. Yasuda-Masuoka, S. D. Kwon and J. S. Yoon, "*Foundry platform technology from high-performance to low-power for new High-Performance Computing (HPC) and IoT era*", 2019 Electron Devices Technol. and Manuf. Conf. (EDTM), Singapore, Singapore, 2019, pp. 1-3.

[5] N. Planes et al., "*28nm FDSOI technology platform for high-speed low-voltage digital applications*", 2012 Symp. VLSI Technol. (VLSIT), Honolulu, HI, 2012, pp. 133-134.

[6] J. Hoentschel, L. Pirro, R. Carter, M. Horstmann, "*22FDX® technologies for ultra-low power IoT, RF and mmWave applications*", Nanoelectronic Devices, **2,** 2019. doi: 10.21494/ISTE.OP.2019.0321.

[7] SAMSUNG, "*Technology offerings (300mm)*". [Online]. Available: https://www.samsungfoundry.com/foundry/homepage/anonymous/technology300nm.do?_mainLayOut=homepageLayout&menuIndex=0201, Accessed on Jul. 2019.

[8] P. Chevalier et al., "*SiGe BiCMOS current status and future trends in Europe*", 2018 IEEE BiCMOS Compound Semicond. Integr. Circuits Technol. Symp. (BCICTS), San Diego, CA, 2018, pp. 64-71.

[9] P. Chevalier et al., "*A 55 nm triple gate oxide 9 metal layers SiGe BiCMOS technology featuring 320 GHz $f_T$/ 370 GHz $f_{max}$ HBT and high-Q millimeter-wave passives*", 2014 IEEE Int. Electron Devices Meeting, San Francisco, CA, 2014, pp. 3.9.1-3.9.3.

[10] X. Wang et al., "*Millimeter wave communication: a comprehensive survey*", in IEEE Commun. Surveys Tut., **20**(3), pp. 1616-1653, 2018.

[11] L. Aluigi, D. Pepe, F. Alimenti and D. Zito, "*K-Band SiGe System-on-Chip Radiometric Receiver for Remote Sensing of the Atmosphere*", IEEE Trans. Circuits Syst. I, Reg. Papers, **64**(12), pp. 3025-3035, 2017.

[12] T. O. Dickson et al., "*The invariance of characteristic current densities in nanoscale MOSFETs and its impact on algorithmic design methodologies and design porting of Si(Ge) (Bi)CMOS high-speed building blocks*", IEEE J. Solid-State Circuits, **41**(8), pp. 1830-1845, 2006.

[13] S. N. Ong et al., "*A 22nm FDSOI technology optimized for RF/mmWave applications*", 2018 IEEE Radio Freq. Integr. Circuits Symp. (RFIC), Philadelphia, PA, 2018, pp. 72-75.

[14] V. P. Trivedi et al., "*A 90nm BiCMOS technology featuring 400GHz $f_{MAX}$ SiGe:C HBT*", 2016 IEEE Bipolar/BiCMOS Circuits Technol. Meeting (BCTM), New Brunswick, NJ, 2016, pp. 60-63.

[15] IHP, "*Low-Volume & Multi-Project Service*", [Online]. Available: https://www.ihp-microelectronics.com/en/services/mpw-prototyping/sigec-bicmos-technologies.html, Accessed on Jul. 2019.

[16] S. Amakawa, "*Theory of gain and stability of small-signal amplifiers with lossless reciprocal feedback*", 2014 Asia-Pacific Microw. Conf., Sendai, Japan, 2014, pp. 1184-1186.

[17] H. Bameri and O. Momeni, "*A high-gain mm-wave amplifier design: an analytical approach to power gain boosting*", IEEE J. Solid-State Circuits, **52**(2), pp. 357-370, 2017.

[18] H. Khatibi, S. Khiyabani and E. Afshari, "*A 173 GHz Amplifier With a 18.5 dB Power Gain in a 130 nm SiGe Process: A Systematic Design of High-Gain Amplifiers Above $f_{max}$*", IEEE Trans. Microw. Theory Techn., **66**(1), pp. 201-214, 2018.

[19] M. Spasaro, F. Alimenti and D. Zito, "*The Theory of Special Noise Invariants*", IEEE Trans. Circuits Syst. I, Reg. Papers, **66**(4), pp. 1305-1318, 2019.

[20] S. P. Voinigescu et al., "*A scalable high-frequency noise model for bipolar transistors with application to optimal transistor sizing for low-noise amplifier design*", IEEE J. Solid-State Circuits, **32**(9), pp. 1430-1439, 1997.

[21] G. Girlando and G. Palmisano, "*Noise figure and impedance matching in RF cascode amplifiers*", IEEE Trans. Circuits Syst. II: Analog Digit. Signal Process., **46**(11), pp. 1388-1396, 1999.

[22] H. A. Haus et al., "*Representation of noise in linear twoports*", Proceedings of the IRE, **48**(1), pp. 69-74, 1960.

[23] H. A. Haus and R. B. Adler, "*Circuit theory of linear noisy networks*", New York, NY, USA: Wiley, 1959.

[24] J. L. Dietrich, "*Unified theory of linear noisy two-ports*", IEEE Trans. Microw. Theory Techn., **61**(11), pp. 3986-3997, 2013.

[25] H. Omran, H. Alahmadi and K. N. Salama, "*Matching properties of femtofarad and sub-femtofarad MOM capacitors*", IEEE Trans. Circuits Syst. I, Reg. Papers, **63**(6), pp. 763-772, 2016.

[26] D. Pepe, I. Chlis and D. Zito, "*Transformer-Based Input Integrated Matching in Cascode Amplifiers: Analytical Proofs*", IEEE Trans. Circuits Syst. I, Reg. Papers, **65**(5), pp. 1495-1504, 2018.

[27] E. Gilbert, "*Impedance matching with lossy components*", IEEE Trans. Circuits and Syst., **22**(2), pp. 96-100, 1975.